\RequirePackage{fix-cm}
\documentclass{svjour3}                     
\smartqed  
\usepackage{graphicx}
\usepackage[british]{babel}
\usepackage{multirow}
\usepackage[caption=false]{subfig}
\usepackage[hypcap=false]{caption}
\usepackage{amssymb}
\usepackage[cmex10]{amsmath}
\usepackage{mathrsfs}
\usepackage{verbatim}
\usepackage{epstopdf}
\usepackage{adjustbox}
\usepackage{color}
\usepackage{makecell}
\usepackage{bm}
\usepackage{hyperref}
\usepackage{enumitem}
\usepackage{times}
\usepackage{setspace}
\usepackage[figuresright]{rotating}
\usepackage{algorithm}
\usepackage{algorithmic}

\usepackage[top=2.5cm, bottom=2.5cm, left=2.5cm, right=2.5cm]{geometry}
\usepackage{appendix}
\usepackage{lineno}
\usepackage{array}
\usepackage{longtable}

\usepackage[square,numbers]{natbib}

\makeatletter
\renewcommand{\@biblabel}[1]{[#1]}
\makeatother

\begin{document}
\setlength{\abovedisplayskip}{4pt}
\setlength{\belowdisplayskip}{4pt}

\title{Nonlocal Adaptive Direction-Guided Structure Tensor Total Variation For Image Recovery \thanks{This work was supported by The Scientific and Technological Research Council of Turkey (TUBITAK) under 115R285.}
}


\author{Ezgi~Demircan-Tureyen \and
        Mustafa E. Kamasak
}


\institute{E. Demircan-Tureyen \at
              Computer Engineering, Istanbul Kultur University, 34158 \\ Computer Engineering, Istanbul Technical University, 34467 \\ Istanbul, Turkey \\
              Tel.: +90-212-4984719\\
              \email{e.demircan@iku.edu.tr}           
           \and
           M. E. Kamasak \at
           Computer Engineering, Istanbul Technical University, 34467 Istanbul, Turkey\\
          \email{kamasak@itu.edu.tr} 
}

\date{Received: date / Accepted: date}

\maketitle

\begin{abstract}
A common strategy in variational image recovery is utilizing the nonlocal self-similarity (NSS) property, when designing energy functionals. One such contribution is nonlocal structure tensor total variation (NLSTV), which lies at the core of this study. This paper is concerned with boosting the NLSTV regularization term through the use of directional priors. More specifically, NLSTV is leveraged so that, at each image point, it gains more sensitivity in the direction that is presumed to have the minimum local variation. The actual difficulty here is capturing this directional information from the corrupted image. In this regard, we propose a method that employs anisotropic Gaussian kernels to estimate directional features to be later used by our proposed model. The experiments validate that our entire two-stage framework achieves better results than the NLSTV model and two other competing local models, in terms of visual and quantitative evaluation.  
\keywords{Directional total variation \and Image recovery \and Nonlocal regularization \and Orientation field estimation \and Structure tensor.}
\end{abstract}

\section{Introduction}
\label{intro}
The general inverse imaging problems seek the recovery of the underlying image $\textbf{f} \in \mathbb{R}^{NC}$ (assuming that each $N$-pixel channel is vectorized and all these $C$ channels are stacked together into a single vector) from an observation model of the form $\textbf{g} = \textbf{H}\textbf{f} + \boldsymbol{\eta}$, where $\boldsymbol{\eta}$ is the noise and $\textbf{H}$ is a known linear operator. The type of the recovery problem is determined by the almost always ill-conditioned $\textbf{H}$, which mainly corresponds to the impulse response of the imaging device. If $\textbf{H} = \textbf{I}$, then the problem is denoising. Besides, it may be a matrix that functions as a mask (i.e., inpainting), or a circulant matrix with all shifts of a blurring kernel (i.e., deblurring). It may also be a composite operator, i.e., $\textbf{H} = \textbf{SF}$, where $\textbf{F}$ blurs and $\textbf{S}$ down-samples (i.e., single-image super-resolution problem), or $\textbf{F}$ applies the Fourier transform and $\textbf{S}$ works as a mask that retains only a subset of the Fourier coefficients (i.e., reconstruction from sparse Fourier measurements). 

The variational approach to such recovery tasks aims at minimizing an energy function of the form:
\begin{equation}
\label{eq:energy} 
E(\textbf{f}) = \frac{1}{2} \Vert \textbf{g} - \textbf{H}\textbf{f}\Vert_2^2  + \tau R(\textbf{f})
\end{equation}
where the former term is known as data fidelity and the latter one is the regularization term that encodes a ‘prior’. The parameter $\tau$ is the regularization parameter that balances the contribution of each term to the cost. The regularization criteria has a critical role in model-driven recovery, thus its proper selection has been immensely studied. One old (but still functional with its extended versions) regularizer is total variation (TV) by \cite{rudin1992nonlinear}, which favors piecewise-constant solutions. For years, TV has been extended in many aspects. Some researchers brought anisotropicity to better handle angled boundaries \cite{grasmair2010anisotropic, lou2015weighted, bayram2012directional, lefkimmiatis2015structure}, some designed higher-order functionals to overcome staircase artifacts by favoring piecewise-smoothness \cite{chan2000high, bredies2010total, papafitsoros2014combined}, and some others employed nonlocal priors to benefit inherent nonlocal self similarity (NSS) as well \cite{gilboa2008nonlocal, lefkimmiatis2015nonlocal}. Structure tensor total variation (STV)  \cite{lefkimmiatis2015structure} is one such TV variant, which redesigns TV in semi-local fashion. Rather than penalizing the gradient, it penalizes a summary of the gradients within a local neighborhood. This neighborhood-awareness provides a better representation of the local image variations, and significantly boosts the performance. Recently in \cite{demircan2019direction, demircan2020adaptive}, we have shown that penalizing the summary of the directional gradients within a patch instead of the gradients increases the performance as long as the local directions are accurately estimated. In this paper, we design a nonlocal counterpart to our prior direction-guided regularizer, under the influence of the nonlocal STV (NLSTV \cite{lefkimmiatis2015nonlocal}), so that the long-distance dependencies across the image can also be modelled. We also design an algorithm that employs anisotropic Gaussian kernels and nonlocal structure tensors to estimate the directional parameters that are needed by the proposed regularizer. 
\vspace{-4mm}
\section{Background}
\label{back}
In \cite{lefkimmiatis2015nonlocal}, the nonlocal counterpart of the STV was proposed. The extension has leveraged the STV term, which had already robustly encoded the local structural variation in semi-local fashion, by widening its scope to involve the nonlocal variations as well. This idea takes its source from the fact that a local patch often has similar nonlocal patches accross the image. This NSS prior was first revealed in the form of nonlocal means (NLM) filtering \cite{buades2011non}, which approximates the underlying intensity value at a spatial coordinate $i$ by taking the weighted average of the patches similar to the one around $i$th pixel.
In \cite{gilboa2008nonlocal} and \cite{elmoataz2008nonlocal}, the idea behind NLM was adopted to variational approach to enrich TV based models. They treated images as a graphs, which allowed them to define nonlocal gradients and the interaction of two distant image points became possible. The nonlocal TV (NLTV) term involves the following nonlocal gradient operator:
\begin{equation}
\label{eq:nlgraphgradient}
(\nabla_w\textbf{f})[i] = \sqrt{w(i,j)}(\textbf{f}[j]-\textbf{f}[i])
\end{equation}
$\forall{j} \in \lbrace 1,2,\cdots, N \rbrace$, where $w(i,j)$ is a function that assigns weights between points by considering the pairwise similarities of the intensities and the relative distances. Therefore, the NLTV prior is defined as follows:
\vspace{-4mm}
\begin{equation}
\label{eq:nltv}
\text{\textit{NLTV}}(\textbf{f}) = \sum_{i=1}^{N} \Vert \nabla_w\textbf{f}[i] \Vert_2
\end{equation}
The rational behind NLTV is to penalize the weighted averages of the gradient magnitudes gathered from the similar patches under the assumption that the gradients of the similar patches are also similar. The NLTV regularization term is designed only for the scalar-valued images, i.e., $C=1$. If we turn back to NLSTV \cite{lefkimmiatis2015nonlocal}, pursuing the same idea with NLTV, it involves a nonlocal structure tensor operator primarily defined as
\begin{equation}
\label{eq:nlStructureTensor}
(S_w\textbf{f})[i] = \sum_{j \in \mathcal{N}[i]} w(i,j) (J\textbf{f})[j](J\textbf{f})[j]^T 
\end{equation}
where $\mathcal{N}[i] = \lbrace j: w(i,j) > 0\rbrace$ and $J$ is the Jacobian operator. It was later reformulated as $(S_w\textbf{f})[i] = (J_w\textbf{f})[i] (J_w\textbf{f})[i]^T$ to be able to decompose the NLSTV into linear functionals (See \cite{lefkimmiatis2015structure, lefkimmiatis2015nonlocal}). Here, the operator $J_w:  \mathbb{R}^{NC} \mapsto \mathbb{R}^{N\times 2 \times NC}$ is nonlocal Jacobian of the form:  
\begin{equation}
\label{eq:nlJacob}
(J_w\textbf{f})[i] = [(D_w\textbf{f}^1)[i] \cdots (D_w\textbf{f}^C)[i]] \in \mathbb{R}^{2 \times NC}
\end{equation}
where each superscript referring to a certain channel and $D_w: \mathbb{R}^N \mapsto \mathbb{R}^{N\times2\times N}$ is nonlocal gradient with the definition: 
\begin{equation}
\begin{split}
\label{eq:nlgradient}
(D_w\textbf{f})[i]^c = [&\sqrt{w(i,1)}(\nabla\textbf{f}^c)[1] \cdots \\ & \quad \sqrt{w(i,N)}(\nabla\textbf{f}^c)[N] \in \mathbb{R}^{2 \times N}
\end{split}
\end{equation}
This definition differs from Eq. \eqref{eq:nlgraphgradient} in the sense that it can also exploit the underlying structure as opposed to the graph-based approach. Having defined the nonlocal Jacobian, the NLSTV is formulated as
\begin{equation}
\small
\label{eq:nlstv}
\text{\textit{NLSTV}}(\textbf{f}) = \sum_{i=1}^N \Vert (J_w \textbf{f}) [i]\Vert_{\mathcal{S}_p} = \Vert J_w \textbf{f} \Vert_{1,p}
\end{equation}
where $\mathcal{S}_p$ is Schatten norm of order $p$, i.e., $\Vert \textbf{X} \Vert_{\mathcal{S}_p} = \Vert \sigma(\textbf{X})\Vert_p$ for a generic matrix \textbf{X} and $\sigma(\cdot)$ returning the singular values \cite{bhatia2013matrix}.
\vspace{-0.3cm}
\section{Method}
\label{method}
Our recovery algorithm involves two stages: (1) the stage that estimates the directional parameters required by the subsequent stage, (2) the stage that solves the inverse problem at hand by employing the proposed regularizer. 
The following subsection starts with the explanation of the proposed regularizer. Since we used the term adaptive direction-guided STV (ADSTV) in \cite{demircan2020adaptive} to refer the extended version of DSTV \cite{demircan2019direction}, which gained the ability of handling multi-directional images, the regularization term that we propose here will be called as nonlocal ADSTV (NLADSTV).
\vspace{-3mm}
\subsection{NLADSTV Regularizer}
From the NLSTV's point of view, we can also further extend our ADSTV to NLADSTV by pursuing the assumption that the directional gradients of the similar patches are also similar. In this respect, we define a \textit{directional nonlocal gradient} that involves directional parameters estimated beforehand, i.e., 
\begin{equation}
\label{eq:nlDirectionalGrad}
\begin{split}
(\tilde{D}^{(\boldsymbol{\alpha},\boldsymbol{\theta})}_w\textbf{f}^c)[i] & =  [\sqrt{w(i,1)}(\tilde{\Pi}_{(\boldsymbol{\alpha},\boldsymbol{\theta})}\nabla\textbf{f}^c)[1] \cdots \\ & \sqrt{w(i,N)}(\tilde{\Pi}_{(\boldsymbol{\alpha},\boldsymbol{\theta})}\nabla\textbf{f}^c)[N]] \in \mathbb{R}^{2 \times N}
\end{split}
\end{equation}
where $(\tilde{\Pi}_{(\boldsymbol{\alpha},\boldsymbol{\theta})}\nabla \textbf{f}^c)[i] = \tilde{\boldsymbol{\Lambda}}_{(\alpha^+,\boldsymbol{\alpha}^\textbf{-}[i])} \textbf{R}_{\boldsymbol{-\theta}[i]}(\nabla \textbf{f}^c)[i]$ for the rotation and scale matrices, i.e.,
\begin{equation}
\small
\label{eq:RandLambda}
{{\textbf{R}}}_{\beta} = \left[ \arraycolsep=1.4pt\def\arraystretch{1.2} \begin{array}{cc}
\cos{\beta} & -\sin{\beta} \\
\sin{\beta} & \cos{\beta}
\end{array} \right] \quad \tilde{\boldsymbol{\Lambda}}_{(\alpha^+,\boldsymbol{\alpha}^\textbf{-}[i])} = \left[\arraycolsep=1.2pt\def\arraystretch{1.2} \begin{array}{cc}
\alpha^+ & 0 \\
0 & \boldsymbol{\alpha}^\textbf{-}[i]
\end{array} \right]
\end{equation}
with $\boldsymbol{\theta} \in [0, \pi)^{N}$, $\boldsymbol{\alpha}^- \in [1, \alpha^+]^{N}$, and $\alpha^+ > 1$. The estimation of these parameters is explained in Section III. B. Thus the \textit{directional nonlocal Jacobian} takes the shape of:
\begin{equation}
\label{eq:nlDirectionalJacob}
\begin{split}
\small
(\tilde{J}^{(\boldsymbol{\alpha},\boldsymbol{\theta})}_w\textbf{f})[i] = [(\tilde{D}^{(\boldsymbol{\alpha},\boldsymbol{\theta})}_w & \textbf{f}^1)[i] \cdots \\ & (\tilde{D}^{(\boldsymbol{\alpha},\boldsymbol{\theta})}_w\textbf{f}^C)[i]] \in \mathbb{R}^{{2 \times NC}}
\end{split}
\end{equation}
and our NLADSTV is nothing but the mixed $\ell_1-S_p$ norm of Eq.\eqref{eq:nlDirectionalJacob}, i.e.,
\begin{equation}
\small
\label{eq:nladstv}
\text{\textit{NLADSTV}}(\textbf{f}) = \sum_{i = 1}^N \Vert (\tilde{J}^{(\boldsymbol{\alpha},\boldsymbol{\theta})}_w \textbf{f})[i] \Vert_{\mathcal{S}_p} = \Vert \tilde{J}^{(\boldsymbol{\alpha},\boldsymbol{\theta})}_w \textbf{f} \Vert_{1,p}
\end{equation}
Since our NLADSTV preserves NLSTV's convexity, the numerical optimization of NLADSTV based Eq. \eqref{eq:energy} is performed by applying the same steps of NLSTV model as described in \cite{lefkimmiatis2015nonlocal}. It was solved by employing the augmented Lagrangian \cite{bertsekas2014constrained} convex optimization tool. Due to its straightforwardness and to avoid repetition, we skip the numerical optimization part in this paper by referring the reader to \cite{lefkimmiatis2015nonlocal}. However, since the adjoint of the directional nonlocal Jacobian is needed by the optimization procedure, adopting Proposition 1 in \cite{lefkimmiatis2015nonlocal}, it is expressed as
\begin{equation}
\small
\label{eq:directionalPatchJacobAdj} 
\begin{split}
(\tilde{J}^{(\boldsymbol{\alpha},\boldsymbol{\theta})^*}_w \textbf{X}) [i] =  [(\tilde{D}^{{(\boldsymbol{\alpha},\boldsymbol{\theta})}^*}_w & \textbf{X}^1)[i] \cdots \\ & (\tilde{D}^{{(\boldsymbol{\alpha},\boldsymbol{\theta})}^*}_w\textbf{X}^C)[i]] \in \mathbb{R}^{C}
\end{split}
\end{equation}
where $\textbf{X} \in \mathbb{R}^{N \times 2 \times NC}$ is an arbitrary matrix field and the adjoint NL gradient acts on $\textbf{X}^c \in \mathbb{R}^{N \times 2 \times N}$ as follows:
\begin{equation}
\small
\label{eq:directionalPatchGradAdj} 
\begin{split}
(\tilde{D}^{{(\boldsymbol{\alpha},\boldsymbol{\theta})}^*}_w\textbf{X}^c)[i] = -{\text{div}} \sum_{j=1}^{N} \Big (\tilde{\Pi}_{(\boldsymbol{\alpha}[i],\boldsymbol{\theta}[i])}^T \sqrt{w(j,i)} \textbf{X}^c[j,:,i] \Big )
\end{split}
\end{equation}
where $\textbf{X}^c[j,:,i] \in \mathbb{R}^2$ refers to the $i$th column of the $j$th matrix of the matrix field $\textbf{X}^c$, div is the discrete divergence operator, and $\tilde{\Pi}_{(\boldsymbol{\alpha}[i],\boldsymbol{\theta}[i])}^T = R_{\boldsymbol{\theta}[i]} \tilde{\Lambda}_{(\alpha^+,\boldsymbol{\alpha}^\textbf{-}[i])}$.

When it comes to the nonlocal weighs, similar to the NLM, the following function is used:
\begin{equation}
\label{eq:nlweights} 
w (i,j) = e^{-d_\rho(i,j)/\beta^2}
\end{equation}
where $d_\rho(\cdot)$ is the distance function of patches and $\beta$ is filtering parameter. Note that, in practical realization of NLSTV, the target space of the nonlocal gradient was shrunk to $\mathbb{R}^{N \times 2 \times L}$, where $L << N$, by using a sparse version of weighting strategy. This strategy seeks the similar patches inside a search window rather than the entire image. Next, it takes only $L$ most similar pair of patches into consideration \cite{lefkimmiatis2015nonlocal}. The distance function that serves this purpose is given below:
\begin{equation}
\label{eq:distance} 
d_\rho (i,j) = \sum_{l = -s/2}^{s/2} G_\rho[l] \vert \textbf{f}[i-l] - \textbf{f}[i+ d(i,j) - l]\vert^2
\end{equation}
where $G_\rho$ is a symmetric weighting kernel of size $s \times s$ and $-r/2 \leq d(i,j) \leq r/2$ returns the relative distance between the pixels $i$ and $j$. The same realization of the weighting function is used by our NLADSTV regularizer.
\vspace{-3mm}
\subsection{Parameter Estimation}
Our regularization term requires three additional parameters at each spatial image point: $\boldsymbol{\theta}[i]$, $\boldsymbol{\alpha}^-[i]$, and a constant $\alpha^+$. The parameter $\boldsymbol{\theta}$ corresponds to the orientation map of the image, while $\boldsymbol{\alpha}^-$ and $\alpha^+$ together determines the anisotropic behaviour of our regularizer. If at a point $i$,  $\boldsymbol{\alpha}^-[i]$ is equal to $\alpha^+$, then the regularizer behaves like NLSTV. Otherwise, as $\boldsymbol{\alpha}^-[i]$ gets closer to one, it gains more anisotropic behaviour, in other words; the sensitivity to the changes towards $\boldsymbol{\theta}[i]$ increases.

\begin{figure*}[t!]
\captionsetup[subfigure]{font = small, labelformat=empty}
 \centering
  \includegraphics[width=\textwidth,keepaspectratio=true]{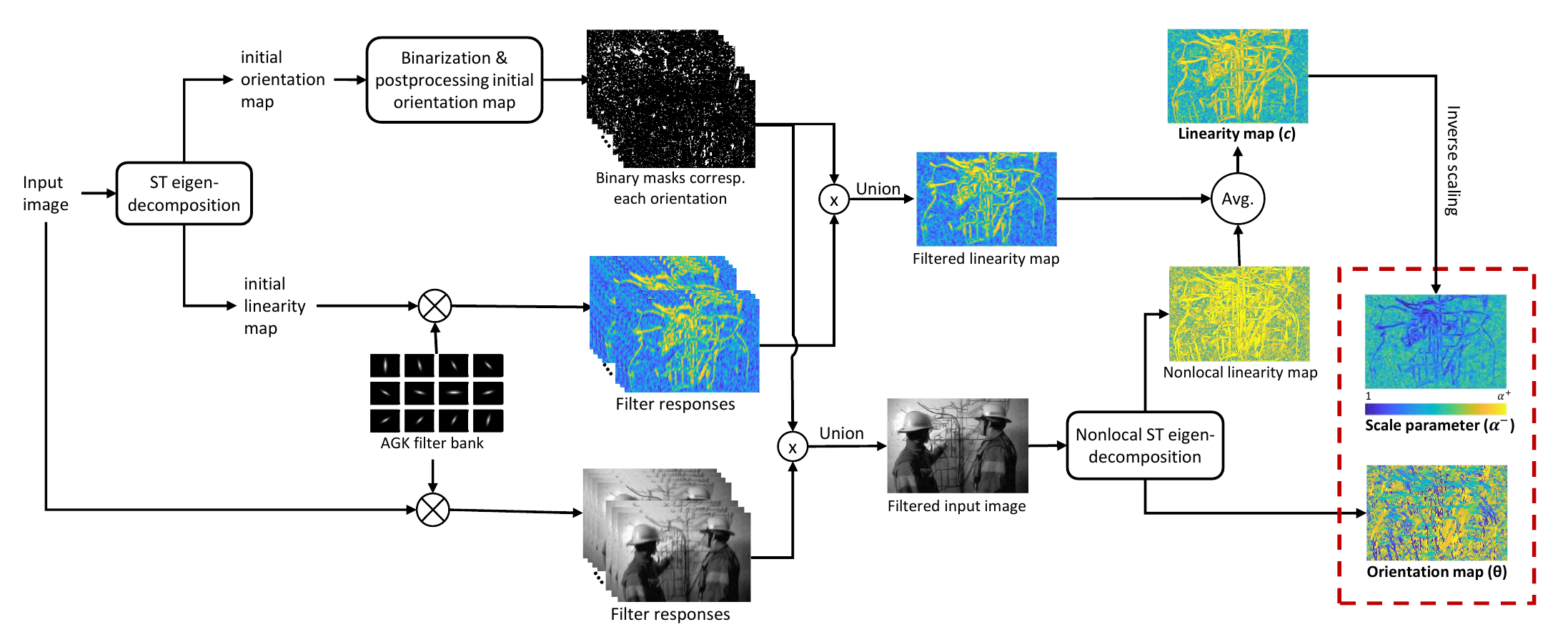}\\
\caption{The flowchart of our DPE procedure for the estimation of directional parameters.}
\label{fig:preprocessor} 
\end{figure*}

In \cite{demircan2020adaptive}, we suggested an algorithm to estimate these parameters. To put it simply, it was applying successive eigendecomposition on the structure tensor (ST) of the observed image (or the luminance information of the observed image for vector-valued images), and TV based smoothing on the parameter fields obtained at multiple scales. In this paper, in order to boost the parameter estimation performance, we propose a different procedure that employs anisotropic Gaussian kernels and nonlocal structure tensor. This subsection elaborates our directional parameter estimation procedure (hereinafter referred to as DPE) by dividing it into steps (See Fig. \ref{fig:preprocessor} for the illustration). Note that, the term ``input image" refers to the luminance information of the observed image, unless otherwise stated. We denote this image as $\tilde{\textbf{g}}$. 

\vspace{-1mm}
\subsubsection{Structure Tensor Eigendecomposition}
We start by obtaining two initial maps through the eigendecomposition of the structure tensor (ST) of the input image. The initial orientation map $\boldsymbol{\theta}^0$ is a map that is composed of the entries $\boldsymbol{\theta}^0[i] \angle \textbf{v}^\textbf{-}[i]$, where $\textbf{v}^\textbf{-}[i]$ is the eigenvector associated with the smallest eigenvalue $\boldsymbol{\lambda}^\textbf{-}[i]$ at an image point $i$. The initial linearity map $\textbf{c}^0$, on the other side, is made of the entries computed as the difference of eigenvalues normalized by the largest eigenvalue $\boldsymbol{\lambda}^\textbf{+}[i]$, i.e., $\textbf{c}^0[i] = {(\boldsymbol{\lambda}^\textbf{+}[i] - \boldsymbol{\lambda}^\textbf{-}[i])}/{\boldsymbol{\lambda}^\textbf{+}[i]}$. This non-negative ratio serves as a measure of local linearity. 
\vspace{-1mm}
\subsubsection{Postprocessing Initial Orientation Map}
The initial orientation map is converted into binary masks each of which is filtering out the orientations except a certain (main) orientation and the orientations falling within the interval $[-\Delta \theta, +\Delta \theta)$ around it. In our work, the main orientations are sampled at 15 degree intervals between $[0, 180)$, thus $\Delta \theta = 7.5$ degree.

Each orientation mask is further processed by using morphological closings in order to fill in the cracks and the holes, so that these filtered-out pixels are also classified by the corresponding orientation. The cracks along the corresponding orientation are closed by using oriented line segments of 5-pixel length as structuring element (SE). The orientation of the line segment matches with the main orientation of the mask. The holes, on the other hand, are closed by using 3-by-3 square SE.   
\vspace{-1mm}
\subsubsection{Filtering with Anisotropic Gaussian Kernels}
Anisotropic Gaussian kernels (AGK) and their derivatives have usually been applied to the edge \cite{zhang2017noise, wang2019multiscale}, line \cite{wang2019noise}, and corner \cite{zhang2020corner} detection tasks. These kernels are nothing but the elongated versions of the isotropic counterpart and defined as  
\begin{equation}
\scriptstyle
\label{eq:AGK}
g(\textbf{g}_L; \sigma, \psi, \beta) = \frac{1}{2\pi\sigma^2}\operatorname{exp}\Bigg( - \frac{1}{2 \sigma^2} \textbf{g}_L^T \textbf{R}_{\beta}\left[ \arraycolsep=1.4pt\def\arraystretch{1.2} \begin{array}{cc}
\scriptstyle
\psi^2 & 0 \\
0 & \scriptstyle \psi^{-2}
\end{array} \right] \textbf{R}_{\beta}^T\textbf{g}_L\Bigg)
\end{equation}
where $\textbf{R}_{\beta}$ is the rotation matrix given in Eq. \eqref{eq:RandLambda}, $\sigma$ represents the scale, $\psi > 1$ denotes the factor of anisotropy ($\psi = 1$ for isotropic kernel), and $\beta$ gives the direction of the kernel. 

We convolve both the input image $\textbf{g}_L$ and the initial linearity map $\textbf{c}^0$ with a filter bank of discrete AGKs covering the main orientations. As mentioned above, the main orientations are sampled at 15 degree intervals between $[0, 180)$. We also fixed the scale parameter as $\sigma = 0.75$ and the anisotropy factor as $\psi = 4$. Among all orientations, we select the ones coinciding with the angles in the postprocessed version of the initial orientation map by performing element-wise multiplication of each filter response with the corresponding mask and by taking the union of those multiplications. Let us call the resulting filtered versions of $\tilde{\textbf{g}}$ and $\textbf{c}^0$ as filtered input image $\tilde{\textbf{g}}_{agk}$ and filtered linearity map $\textbf{c}_{agk}$, respectively.
\vspace{-3mm}
\subsubsection{Nonlocal Structure Tensor Eigendecomposition}
This step is very similar to the first step, except this time the structure tensor is nonlocal and computed on the filtered input image $\tilde{\textbf{g}}_{agk}$, rather than $\tilde{\textbf{g}}$. The rational behind preferring not to apply the nonlocal ST eigendecomposition in the first step is due to its less robustness to the noise or artifacts, when compared to its local counterpart. The orientation map $\boldsymbol{\theta}$ with the entries $\boldsymbol{\theta}[i] \angle \textbf{v}_{nl}^\textbf{-}[i]$, where the subscript $nl$ indicates that the eigenvector is derived from the nonlocal ST, is the final orientation map now, which will be fed into the restoration process as it is. On the other side, the entries of the nonlocal linearity map is obtained as $\textbf{c}_{nl}[i] = {(\boldsymbol{\lambda}_{nl}^\textbf{+}[i] - \boldsymbol{\lambda}_{nl}^\textbf{-}[i])}/{\boldsymbol{\lambda}_{nl}^\textbf{+}[i]}$ and averaged with the filtered linearity map to reach the final linearity map, i.e., $\textbf{c} = (\textbf{c}_{agk} + \textbf{c}_{nl}) / 2$. 
\vspace{-3mm}
\subsubsection{The Final Directional Parameters}
The orientation map $\boldsymbol{\theta}$ to be used by the NLADSTV regularizer is the direct output of nonlocal ST eigendecomposition as stated in the previous step. The parameter $\boldsymbol{\alpha}^-$, on the other hand, is computed by inversely scaling the entries of $\textbf{c} \in \mathbb{R}^N$ onto the range $[1, \alpha^+]$, i.e.,
\begin{equation}
\small
\label{eq:alphaFinal}
\boldsymbol{\alpha}^-[i] = \frac{\alpha^+ - 1}{\text{max}({\textbf{c}})-\text{min}({\textbf{c}})} (\text{max}({\textbf{c}}) - {\textbf{c}}[i]) + 1
\end{equation}

In Fig. \ref{fig:preprocessor} dashed red box shows the final directional parameters.

\vspace{-3mm} 
\subsection{Overall Algorithm}
Our recovery framework starts by estimating the directional parameters as described in Subsection B. Next, the NLADSTV regularized inverse problem (See Eq. \eqref{eq:minimizationprob}) is solved by following the optimization scheme in \cite{lefkimmiatis2015nonlocal}. The overall algorithm is provided in Algorithm 1. This algorithm actually repeats the steps in the Algorithm 1 in \cite{lefkimmiatis2015nonlocal}, except the presence of the DPE procedure (called at third line) and the nonlocal Jacobian operator switched to our directional nonlocal Jacobian. 

As mentioned earlier, we don't go into the details of the numerical optimization in this paper, however the Algorithm 1 needs some explanations. It solves the following NLADSTV regularized inverse problem: 
\begin{equation}
\label{eq:minimizationprob} 
\hat{\textbf{{f}}} = \underset{\textbf{f}}{\operatorname{argmin}} \ \frac{1}{2} \Vert \textbf{g} - \textbf{H}\textbf{f}\Vert_2^2  + \tau \Vert \tilde{J}^{(\boldsymbol{\alpha},\boldsymbol{\theta})}_w \textbf{f} \Vert_{1,p} + \iota_\mathcal{C}(\textbf{f})
\end{equation}
where $\iota_\mathcal{C}$ is the indicator function of a convex set $\mathcal{C}$ that takes the value 0 if $\textbf{f} \in \mathcal{C}$ and $\infty$ otherwise. The set $\mathcal{C}$ corresponds to an additional constraint, such as nonnegativity, and if there is no constraint, then $\mathcal{C} = \mathbb{R}^{NC}$. The authors of \cite{lefkimmiatis2015nonlocal} reformulates this problem in constrained form by using auxiliary variables and seeks a solution through augmented Lagrangian methods \cite{bertsekas2014constrained}. They employ alternating-direction method of multipliers (ADMM) \cite{eckstein1992douglas}.

In the line 5, the $\operatorname{prox}$ operator at the RHS corresponds to the proximal map of the mixed $\ell_1-S_p$ norm weighted by $\tau/\mu$. For a function $h$ evaluated at $z$, it is defined as:
\begin{equation}
\label{eq:prox} 
\operatorname{prox}_{h}(\textbf{z}) = \underset{\textbf{f}}{\operatorname{argmin}} \ \frac{1}{2} \Vert \textbf{f} - \textbf{z} \Vert^2 + h(\textbf{f})
\end{equation}
While deriving a minimization approach in \cite{lefkimmiatis2015nonlocal} for the NLSTV regularized inverse problem, Eq. \eqref{eq:prox} is reduced to the proximal map of $\mathcal{S}_p$ norm by inputting the entries $\Omega[i] \in \mathbb{R}^{2 \times LC}$ of $\Omega = \tilde{J}^{(\boldsymbol{\alpha},\boldsymbol{\theta})}_w \textbf{f}^{(t)} + \textbf{s}^t_1$ separately. Next, through the SVD decomposition of $\Omega[i]$, the problem is reduced to the proximal map of $\ell_p$ norm with an input consisting of the singular values of $\Omega[i]$. For the efficient evaluation of $\operatorname{prox}_{p}$, they utilize iterative proximal algorithm proposed in \cite{liu2010efficient}. (See Section IV in \cite{lefkimmiatis2015nonlocal} for the details).

In the line 6, $P_{\mathcal{C}}$ is nothing but the projection of $\textbf{f}^{(t)} + \textbf{s}_2^{(t)}$ onto the convex set $\mathcal{C}$. It is defined as $P_{\mathcal{C}} = \min(\max(0,\textbf{f}),\gamma)$ for the set $\mathcal{C} = \{\textbf{f} \in \mathbb{R}^{NC} : 0 \leq \textbf{f}[i] \leq \gamma, \forall i = 1, \cdots, NC \}$ that we consider \cite{lefkimmiatis2015nonlocal}.

The rest of the Algorithm 1 keeps realizing the steps of ADMM iterations and updating the Lagrange multipliers (The reader is referred to \cite{lefkimmiatis2015nonlocal} for details).

\begin{algorithm}[h!]
\caption{Algorithm for NLADSTV-based recovery (adopted from \cite{lefkimmiatis2015nonlocal})}\label{euclid}
\begin{spacing}{1.1}
\begin{algorithmic}[1]
\small
\STATE \textbf{INPUT:} $\textbf{g}$, $\textbf{H}$, $\tau > 0$, $\mu > 0$, $p \geq 1$, ${\alpha}^+ > 1$ 
\STATE \textbf{INIT:} ${\textbf{f}}^{(0)} = \textbf{g}$, \quad $\textbf{s}^{(0)}_1 = \textbf{0}$, \quad $\textbf{s}^{(0)}_2 = \textbf{0}$, \quad $t = 0$, 
\STATE $(\boldsymbol{\theta}, \boldsymbol{\alpha^-}) \leftarrow $ \textbf{DPE}$(\textbf{g}, \alpha^+)$ 
\WHILE{stopping criterion is not satisfied}
\STATE $\textbf{z}_1^{(t+1)} \leftarrow \operatorname{prox}_{\frac{\tau}{\mu}\Vert \cdot \Vert_{1,p}}(\tilde{J}^{(\boldsymbol{\alpha},\boldsymbol{\theta})}_w \textbf{f}^{(t)} + \textbf{s}^t_1)$
\STATE $\textbf{z}_2^{(t+1)} \leftarrow P_{\mathcal{C}}(\textbf{f}^{(t)} + \textbf{s}_2^{(t)})$
\STATE $\textbf{B} \leftarrow \big( \frac{1}{\mu} \textbf{H}^T \textbf{H} + \tilde{J}^{(\boldsymbol{\alpha},\boldsymbol{\theta})}_{w^*} \tilde{J}^{(\boldsymbol{\alpha},\boldsymbol{\theta})}_w + \textbf{I}\big)$
\STATE $\boldsymbol{\omega}_1^{(t+1)} \leftarrow \textbf{z}_1^{(t+1)} - \textbf{s}^{(t)}_1$
\STATE $\boldsymbol{\omega}_2^{(t+1)} \leftarrow \textbf{z}_2^{(t+1)} - \textbf{s}^{(t)}_2$
\STATE $\textbf{f}^{(t+1)} \leftarrow \textbf{B}^{-1} \big( \frac{1}{\mu} \textbf{H}^T \textbf{g} + \tilde{J}^{(\boldsymbol{\alpha},\boldsymbol{\theta})}_{w^*} \boldsymbol{\omega}_1^{(t+1)} + \boldsymbol{\omega}_2^{(t+1)} \big)$
\STATE $\textbf{s}^{(t+1)}_1 \leftarrow \textbf{s}^{(t)}_1 +  \tilde{J}^{(\boldsymbol{\alpha},\boldsymbol{\theta})}_w \textbf{f}^{(t+1)} - \textbf{z}_1^{(t+1)}$
\STATE $\textbf{s}^{(t+1)}_2 \leftarrow \textbf{s}^{(t)}_2 + \textbf{f}^{(t+1)} - \textbf{z}_2^{(t+1)}$
\STATE $t \leftarrow t+1$
\ENDWHILE
\end{algorithmic}
\end{spacing}
\end{algorithm}
\vspace{-0.3cm}
\section{Experimental Results}
\label{exp_results}
\begin{figure*}[t!]
\newcommand{\mywidth}{1.55cm}
\captionsetup[subfigure]{font = small, labelformat=empty}
 \centering
  \includegraphics[width=\textwidth,keepaspectratio=true]{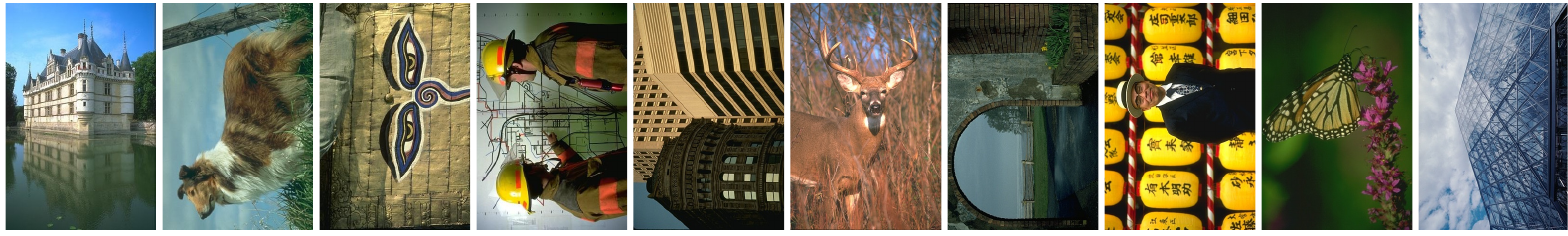}
  \vspace{-0.4cm}
\caption{Thumbnails of the grayscale and color images used in the experiments. From left to right and top to bottom: \textit{Chateau, Dog, Mural, Workers, Buildings, Deer, Arch, Man, Monarch}, and \textit{Structure}.}
\label{fig:images} 
\end{figure*}
We consider the problems of image denoising and image deblurring by confronting four related methods: STV, ADSTV, NLSTV, and NLADSTV. The experiments are conducted on the vector-valued images shown in Fig. \ref{fig:images}. The images are taken from the popular Berkeley Segmentation Dataset (BSD) \cite{arbelaez2010contour}. For the quantitative evaluation, we use the peak signal-to-noise ratio (PSNR). The source codes of STV\footnote{\scriptsize{\url{https://github.com/cig-skoltech/Structure_Tensor_Total_Variation}}} and NLSTV\footnote{\scriptsize{\url{https://github.com/cig-skoltech/NLSTV}}} that we use were made publicly available by the authors on GitHub. Our NLADSTV is implemented on top of NLSTV. 

Since the algorithms under consideration are all aiming to minimize Eq. \eqref{eq:energy}, they need regularization parameter $\tau$ to be tuned. We fine-tuned $\tau$ such that it leads to the best PSNR in all experiments. For both STV and ADSTV, the Gaussian kernel of support $3 \times 3$ pixel with the standard deviation $\sigma = 0.5$ is used. For NLSTV and NLADSTV, on the other hand, the sizes of the neighborhood (used for the nonlocal weights) and the search window are set to $7 \times 7$ and $11 \times 11$, respectively. Also, the number of the similar patches $L = 9$, as suggested in \cite{lefkimmiatis2015nonlocal}. For the sake of fairness, we fixed the free parameter $\alpha^+$ of ADSTV and NLADSTV as 4, but it is possible to increase the restoration quality of these methods by fine-tuning $\alpha^+$. When it comes to the setting of the DPE procedure, the Gaussian kernel used for the computation of the local structure tensor has $9 \times 9$ pixel support and the standard deviation of $1.5$, while the size of the neighborhood is $5 \times 5$, the size of the search window is $11 \times 11$, and $L = 13$. In addition to them, note that the order of the Schatten norm $p=1$ for all regularizers under consideration.
\vspace{-3mm}
\subsection{Image Denoising}
\begin{table*}[t!]
  \centering
  \caption{PSNR Comparisons on Image Denoising}
  \vspace{-2mm}
  \setlength\tabcolsep{1.8pt}
  \Large
    \resizebox{\textwidth}{!}{\begin{tabular}{l|cccc|cccc|cccc|cccc}
    \hline
    \hline
    $\sigma_\eta$:    & \multicolumn{4}{c|}{\textbf{0.05}} & \multicolumn{4}{c|}{\textbf{0.1}} & \multicolumn{4}{c|}{\textbf{0.15}} & \multicolumn{4}{c}{\textbf{0.2}} \\
    \hline
    Method: & \textbf{\ \ STV \ \ } & \textbf{\ ADSTV \ } & \textbf{\ NLSTV \ } & \textbf{NLADSTV} & \textbf{\ \ STV \ \ } & \textbf{\ ADSTV \ } & \textbf{\ NLSTV \ } & \textbf{NLADSTV} & \textbf{\ \ STV \ \ } & \textbf{\ ADSTV \ } & \textbf{\ NLSTV \ } & \textbf{NLADSTV} & \textbf{\ \ STV \ \ } & \textbf{\ ADSTV \ } & \textbf{\ NLSTV \ } & \textbf{NLADSTV} \\
    \hline
    Chateau & 32.38 & 32.73 & 32.92 & \textbf{33.15} & 28.79 & 29.33 & 29.41 & \textbf{29.71} & 26.93 & 27.60 & 27.56 & \textbf{27.93} & 25.74 & 26.37 & 26.29 & \textbf{26.71} \\
    Dog   & 32.19 & 32.56 & 32.25 & \textbf{32.61} & 29.00 & 29.38 & 29.09 & \textbf{29.44} & 27.44 & 27.82 & 27.59 & \textbf{27.87} & 26.46 & 26.78 & 26.63 & \textbf{26.85} \\
    Mural & 30.16 & 30.41 & 30.20 & \textbf{30.46} & 26.59 & 26.95 & 26.74 & \textbf{27.14} & 24.78 & 25.19 & 25.03 & \textbf{25.47} & 23.61 & 24.00 & 23.88 & \textbf{24.35} \\
    Workers & 31.90 & 32.38 & 32.40 & \textbf{32.87} & 28.04 & 28.69 & 28.66 & \textbf{29.35} & 25.94 & 26.65 & 26.60 & \textbf{27.35} & 24.55 & 25.20 & 25.13 & \textbf{25.92} \\
    Buildings & 31.24 & 31.82 & 31.35 & \textbf{32.13} & 27.44 & 28.15 & 27.72 & \textbf{28.55} & 25.47 & 26.25 & 25.90 & \textbf{26.74} & 24.20 & 24.93 & 24.64 & \textbf{25.50} \\
    Deer  & 31.49 & \textbf{31.76} & 31.31 & 31.69 & 28.30 & \textbf{28.65} & 28.22 & 28.54 & 26.69 & \textbf{27.04} & 26.67 & 26.90 & 25.64 & \textbf{25.97} & 25.66 & 25.80 \\
    Arch  & 30.86 & 31.03 & 30.92 & \textbf{31.07} & 27.63 & 27.90 & 27.76 & \textbf{28.14} & 26.08 & 26.38 & 26.27 & \textbf{26.61} & 25.15 & 25.34 & 25.36 & \textbf{25.63} \\
    Man   & 33.42 & 33.64 & 33.65 & \textbf{33.97} & 29.54 & 29.85 & 29.88 & \textbf{30.31} & 27.38 & 27.71 & 27.71 & \textbf{28.24} & 25.90 & 26.19 & 26.16 & \textbf{26.75} \\
    Monarch & 35.16 & 35.52 & 35.42 & \textbf{35.73} & 31.68 & 32.09 & 32.03 & \textbf{32.37} & 29.73 & 30.17 & 30.21 & \textbf{30.51} & 28.39 & 28.78 & 28.88 & \textbf{29.23} \\
    Structure & 30.00 & 30.31 & 30.32 & \textbf{30.65} & 26.27 & 26.68 & 26.58 & \textbf{27.05} & 24.51 & 24.92 & 24.82 & \textbf{25.29} & 23.45 & 23.82 & 23.73 & \textbf{24.18} \\
    \hline
    \hline
    \textbf{Avg.} & 31.88 & 32.22 & 32.07 & \textbf{32.43} & 28.33 & 28.77 & 28.61 & \textbf{29.06} & 26.50 & 26.97 & 26.84 & \textbf{27.29} & 25.31 & 25.74 & 25.64 & \textbf{26.09} \\
    \hline
    \hline
    \end{tabular}}%
  \label{tab:denoising}%
\end{table*}%

We consider additive i.i.d. Gaussian noise with four noise levels $\sigma_\eta = \{ 0.05, 0.1, 0.15, 0.2\}$. Table \ref{tab:denoising} reports the PSNR scores obtained by using four competing methods. By inspecting the results, one observes that the local regularizers perform worse than their nonlocal counterparts, except for the \textit{Deer} image, which may not be exhibiting NSS property. STV is the least performing regularizer for all four noise levels. As was shown in \cite{demircan2019direction}, ADSTV systematically outperforms STV with the improvement around 0.4 dBs. Another observation is, as the noise level increases, ADSTV also produces better results than NLSTV. Our NLADSTV, on the other hand, is the best performing regularizer with 0.4 dBs improvement over NLSTV and 0.3 dBs over its local counterpart ADSTV, on average.

\begin{figure*}[t!]
\newcommand{\mywidthwo}{6.7cm}
\captionsetup[subfigure]{labelformat=empty}
 \centering 
  \includegraphics[width=\textwidth,keepaspectratio=true]{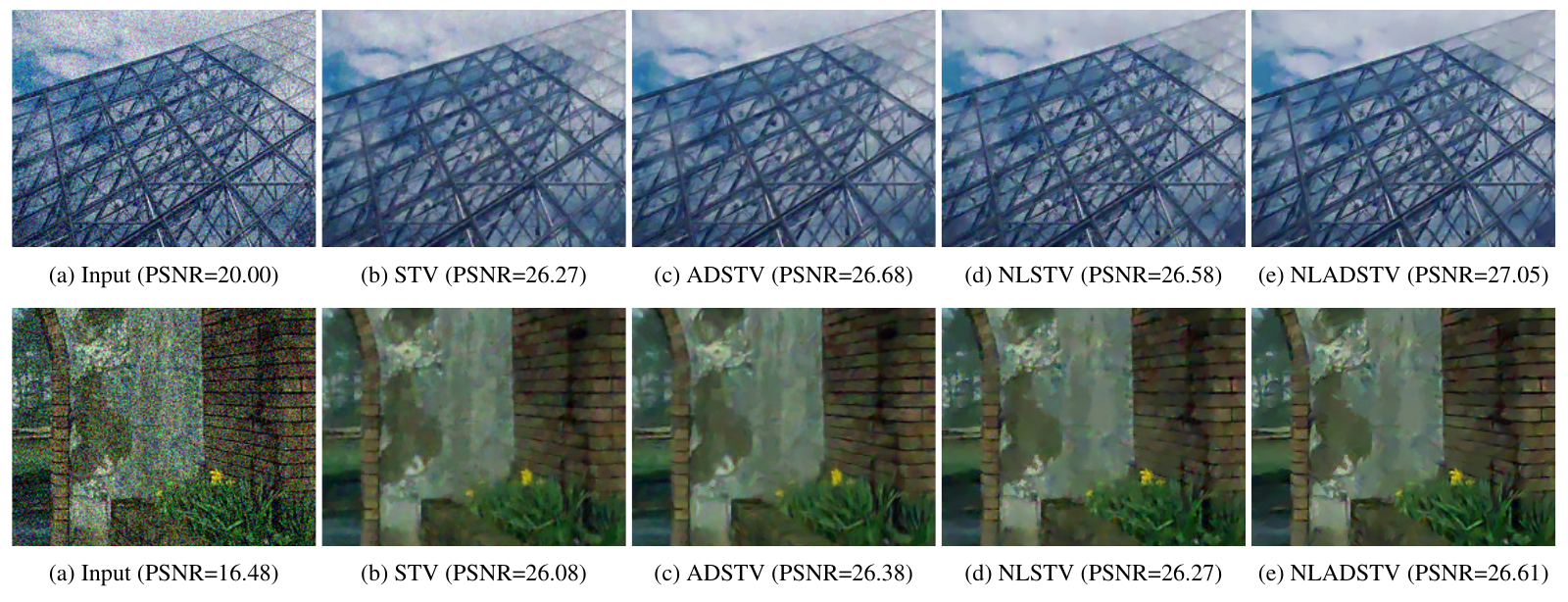}
\vspace{-2mm}
  \caption{\small{The detail patches cropped from restored versions of the noisy \textit{Structure} and \textit{Arch} images. \textit{Structure} image is degraded the noise level of $\sigma_\eta = 0.1$.  \textit{Arch} image is degraded by the noise level of $\sigma_\eta = 0.15$.}}  
 \label{fig:denoising} 
\end{figure*}

Fig. \ref{fig:denoising} is also provided for the visual judgement. We demonstrate two representative detail patches. As we observe from the patch taken from the \textit{Structure} image, the straight lines seem smoother in the NLADSTV reconstruction. The oil-painting-like artifacts present in the STV and NLSTV are far less visible in ADSTV and NLADSTV.  When compared to the ADSTV, the NLADSTV is better at removing noise while preserving details, thanks to its nonlocality. In the \textit{Arch} image, the details of the plant and the bricks around the arch are more distinguishably reconstructed by our NLADSTV than the other methods. 
\vspace{-3mm}
\subsection{Image Deblurring}
In deblurring problem, instead of feeding the direct observation (blurred image) to the DPE procedure, we input the deblurred version of the image by a Wiener filter, since otherwise the blur effect causes the poor localization of edges and the loss of some small gaps between parallel lines. By means of Wiener filter, our DPE procedure can better estimate the directional parameters.
\begin{table*}[t!]
  \centering
  \caption{PSNR Comparisons on Image Deblurring}
  \vspace{-2mm}
  \setlength\tabcolsep{1.5pt}
  \Large
    \resizebox{\textwidth}{!}{\begin{tabular}{l|cccc|cccc|cccc|cccc}
    \hline
    \hline
          & \multicolumn{8}{c|}{\textbf{Gaussian PSF}}                    & \multicolumn{8}{c}{\textbf{Motion PSF}} \\
    \hline
    \textbf{BSNR:} & \multicolumn{4}{c|}{\textbf{20 dB}} & \multicolumn{4}{c|}{\textbf{30 dB}} & \multicolumn{4}{c|}{\textbf{20 dB}} & \multicolumn{4}{c}{\textbf{30 dB}} \\
    \hline
          & \textbf{\ \ STV \ \ } & \textbf{\ ADSTV \ } & \textbf{\ NLSTV \ } & \textbf{NLADSTV} & \textbf{\ \ STV \ \ } & \textbf{\ ADSTV \ } & \textbf{\ NLSTV \ } & \textbf{NLADSTV} & \textbf{\ \ STV \ \ } & \textbf{\ ADSTV \ } & \textbf{\ NLSTV \ } & \textbf{NLADSTV} & \textbf{\ \ STV \ \ } & \textbf{\ ADSTV \ } & \textbf{\ NLSTV \ } & \textbf{NLADSTV} \\
    \hline
    Chateau & 25.62 & 25.92 & 25.83 & \textbf{26.04} & 27.85 & 28.10 & 27.77 & \textbf{28.27} & 27.73 & 28.14 & 28.15 & \textbf{28.43} & 31.81 & 32.28 & 32.44 & \textbf{32.78} \\
    Dog   & 27.18 & \textbf{27.25} & 27.19 & \textbf{27.25} & 28.70 & 28.84 & 28.76 & \textbf{28.90} & 29.54 & 29.76 & 29.59 & \textbf{29.85} & 33.81 & 34.14 & 33.88 & \textbf{34.24} \\
    Mural & 23.91 & 24.13 & 24.08 & \textbf{24.20} & 26.12 & 26.25 & 26.34 & \textbf{26.42} & 26.30 & 26.71 & 26.56 & \textbf{26.93} & 31.17 & 31.51 & 31.34 & \textbf{31.66} \\
    Workers & 23.80 & 24.09 & 24.04 & \textbf{24.37} & 26.26 & 26.52 & 26.62 & \textbf{26.89} & 26.38 & 26.99 & 26.96 & \textbf{27.74} & 31.42 & 31.99 & 32.16 & \textbf{32.78} \\
    Buildings & 24.02 & 24.21 & 24.37 & \textbf{24.58} & 25.98 & 26.25 & 26.24 & \textbf{26.51} & 27.22 & 27.71 & 27.50 & \textbf{28.08} & 31.81 & 32.42 & 31.89 & \textbf{32.65} \\
    Deer  & 27.39 & \textbf{27.54} & 27.30 & 27.43 & 29.28 & \textbf{29.44} & 29.28 & 29.39 & 29.25 & \textbf{29.54} & 29.20 & 29.51 & 33.90 & \textbf{34.20} & 33.81 & 34.19 \\
    Arch  & 25.93 & 26.00 & 26.00 & \textbf{26.11} & 27.58 & 27.58 & 27.63 & \textbf{27.68} & 27.64 & 27.84 & 27.73 & \textbf{28.06} & 31.88 & 32.07 & 31.96 & \textbf{32.27} \\
    Man   & 25.38 & \textbf{25.45} & 25.40 & \textbf{25.45} & 28.15 & 28.28 & 28.35 & \textbf{28.44} & 27.34 & 27.63 & 27.74 & \textbf{28.09} & 32.40 & 32.72 & 32.86 & \textbf{33.23} \\
    Monarch & 30.15 & 30.15 & \textbf{30.39} & 30.31 & 32.54 & 32.63 & \textbf{32.97} & \textbf{32.97} & 33.20 & 33.64 & 33.70 & \textbf{34.00} & 38.40 & 38.76 & 38.88 & \textbf{39.20} \\
    Structure & 22.99 & 23.10 & 23.02 & \textbf{23.14} & 24.22 & 24.35 & 24.30 & \textbf{24.46} & 25.32 & 25.64 & 25.44 & \textbf{25.90} & 29.79 & 30.13 & 30.06 & \textbf{30.43} \\
    \hline
    \hline
    Avg.  & 25.64 & 25.79 & 25.76 & \textbf{25.89} & 27.67 & 27.82 & 27.83 & \textbf{27.99} & 27.99 & 28.36 & 28.26 & \textbf{28.66} & 32.64 & 33.02 & 32.93 & \textbf{33.34} \\
    \hline
    \hline
    \end{tabular}}%
  \label{tab:deblurring}%
\end{table*}%
\begin{figure*}[t!]
\newcommand{\mywidthwo}{6.7cm}
\captionsetup[subfigure]{labelformat=empty}
 \centering 
 \includegraphics[width=\textwidth,keepaspectratio=true]{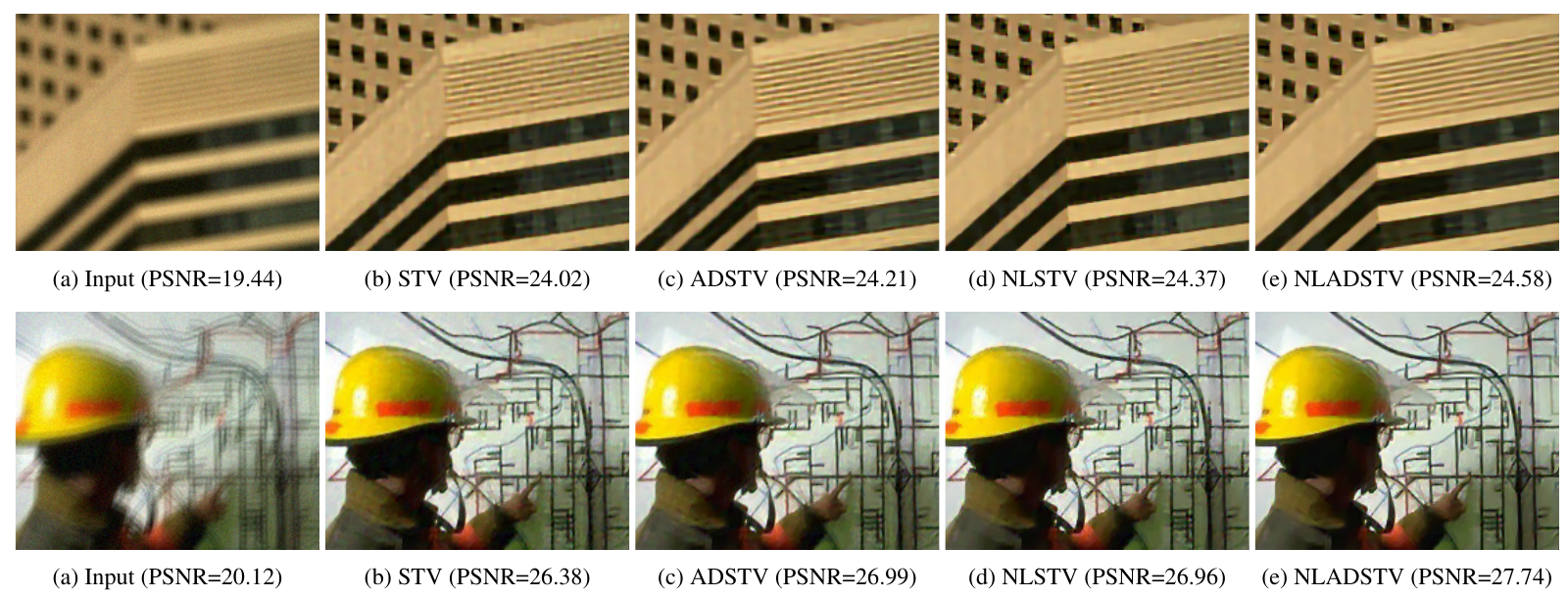}
\vspace{-2mm}
  \caption{\small{The detail patches cropped from restored versions of the blurred and noisy \textit{Buildings} and \textit{Workers} images. \textit{Buildings} image is degraded by Gaussian blur and noise level of BSNR = 20 dBs.  \textit{Workers} image is degraded by motion blur and BSNR = 20 dBs.}}  
 \label{fig:deblurring} 
\end{figure*}

Table \ref{tab:deblurring} reports the deblurring performances of the methods under consideration for two types of blurring kernels. We consider a Gaussian kernel of support $9 \times 9$ pixel with $\sigma_b = 6$ and a motion kernel of support $19 \time 19$ pixel. The blurred images are also degraded with Gaussian noise corresponding to blurred SNR (BSNR = variance of $\textbf{H}\textbf{f}$ divided by $\sigma_\eta^2$) of 20 and 30 dBs. While the deblurring results for motion blur are on par with the denoising results, the improvement that we obtain with NLADSTV is less notable (around 0.15 dBs over both ADSTV and NLSTV on average) when it comes to the Gaussian blur.  

We demonstrate two exemplary results in Fig. \ref{fig:deblurring}. As one can observe from the detail patch of the \textit{Buildings} image, the artifacts between the closely parallel lines are far less visible in the NLADSTV regularized solution. In the \textit{Workers} image again, the the cloudy look around the lines is more apparent in the other methods than in our NLADSTV.
\vspace{-0.3cm}
\section{Discussion}
According to the experiments conducted, our approach seems to produce promising results. However, the effectiveness of the NLADSTV based regularization highly depends on the performance of the directional parameter estimation (DPE) procedure. In the case of Gaussian blur for instance, since our DPE procedure was able to estimate less reliable parameters, the PSNR gain was not as satisfying as it was in the other experiments. Fortunately, because the structure tensor is a positive semi-definite matrix, the linearity factor $\textbf{c}[i]$ also behaves like a coherence factor, which measures the reliability of the estimated orientations. Thus for the unreliable orientation estimations, $\textbf{c}[i]$ becomes small and the NLADSTV starts working like the NLSTV. Therefore, the restoration quality never falls below the one that we obtain by using NLSTV. 

The NLADSTV comes up with additional computational demands. The DPE procedure takes around 10 sec. on average, on a computer equipped with Intel Core Processor i7-7500U (2.70-GHz) with 16 GB of memory. Besides, the convergence of the NLADSTV based problem requires more iterations than the NLSTV, which makes the entire framework run 2 times slower than the NLSTV based image recovery. 
\vspace{-0.3cm}
\section{Conclusion}
This paper proposes a competitive framework for variational image recovery. We establish a two-stage algorithm to advance the nonlocal structure tensor total variation (NLSTV) regularizer, which already exploits local structural regularity and nonlocal self-similarity properties. We further enrich these priors by encoding some spatially varying directional characteristics of the underlying image. In order to understand these characteristics, we design an algorithm that estimates the directional parameters required by our regularizer. The experiments prove that the proposed method achieves significant performance improvements over its local counterpart and the NLSTV. 

\newpage
\bibliographystyle{spbasic}      
\bibliography{template}   

%
%

\end{document}